\documentclass[12pt]{article}
\usepackage{amsmath,amsfonts, feynmp, epsf,multirow}
\usepackage{amssymb}
\usepackage{graphicx}
\usepackage{grffile}
\input epsf

\def\bZ {\mathbb{Z}}

\def\be{\begin{equation}}
\def\ee{\end{equation}}
\def\bea{\begin{eqnarray}}
\def\eea{\end{eqnarray}}
\def\ie{\begin{equation}\begin{aligned}}
\def\fe{\end{aligned}\end{equation}}

\newcommand{\m}{\mu}
\newcommand{\n}{\nu}

\newcommand{\A}{{\alpha}}
\newcommand{\B}{{\beta}}
\newcommand{\C}{{\gamma}}

\makeatletter\@addtoreset{equation}{section}\makeatother

\newcommand{\vev}[1]{{\left< {#1} \right>}}

\newcommand{\tr}{{\rm tr\,}}
\newcommand{\cA}{{\mathcal A}}

\newcommand{\cO}{{\mathcal O}}

\renewcommand{\title}[1]{\vbox{\center\LARGE{#1}}\vspace{5mm}}
\renewcommand{\author}[1]{\vbox{\center#1}\vspace{5mm}}

\makeatletter\@addtoreset{equation}{section}\makeatother

\begin{document}\begin{titlepage}
\begin{center}

\title{\LARGE {\textsc{Non-Minimal Higher-Spin DS$_4$/CFT$_3$}}}
\vspace{0.8cm}
Chi-Ming Chang, Abhishek Pathak and Andrew Strominger

\vspace{1cm}

{\it  Center for the Fundamental Laws of Nature, Harvard University,\\
Cambridge, MA 02138, USA}

%\author{}
\date{\today}
\begin{abstract}
We conjecture that the level $k$ $U(N)$ Chern-Simons theory coupled to free anticommuting scalar matter  in the fundamental is dual to non-minmal higher-spin Vasiliev gravity in dS$_4$ with parity-violating phase $\theta_0={\pi N\over 2k}$ and Neumann boundary conditions for the scalar.  Related conjectures are made for fundamental commuting spinor matter and critical theories. This generalizes a recent conjecture relating  the minimal Type A Vasiliev theory in dS$_4$ to the $Sp(N)$ model with fundamental real anti-commuting scalars.
\end{abstract}
\end{center}
\end{titlepage}

\setcounter{page}{1}
\pagenumbering{arabic}
\tableofcontents

\section{Introduction and Summary }

The dS/CFT conjecture posits that that quantum gravity on de Sitter space (dS) is holographically dual to 
a conformal field theory (CFT) living on the spacelike boundary of dS at future infinity \cite{Strominger:2001pn, Witten:2001kn, Maldacena:2002vr}.  Development of this conjecture was hindered by the absence of any concrete example. Recently, an example was proposed:  Vasiliev type A theory in dS$_4$ is conjecturally dual to the $Sp(N)$ CFT$_3$ with anti-commuting scalars \cite{Anninos:2011ui}. This opened the door to detailed  investigations of the structure of higher-spin dS holography  $e.g.$\cite{Banerjee:2013mca}-\cite{Ng:2012xp}.
% \cite{Banerjee:2013mca,Karch:2013oqa,Das:2013qea,Anninos:2013rza,Jin:2013lqa,Anninos:2012ft,Das:2012dt,Anninos:2012qw,Ng:2012xp}.

   Vasiliev actually constructed families of classical higher-spin gravity theories in dS$_4$ labelled by a parity-violating phase $\theta_0$ and, at the quantum level, a loop counting parameter $g_{dS}^2$. In this paper we generalize the construction of \cite{Anninos:2011ui} and propose duals for this family of theories. They are level $k$ $U(N)$ Chern-Simons theories coupled to wrong-statistics scalars or spinors with $N=g_{dS}^{-2}$ and 't Hooft coupling $\lambda={N \over k}$ a certain real function of $\theta_0$. We rely heavily on the analysis of \cite{Giombi:2012ms,Giombi:2011kc,Aharony:2011jz} in which the analogous question was addressed in AdS$_4$. The dualities are displayed in table 1. 
   
The existence of such dualities was anticipated in \cite{Banerjee:2013mca, Anninos:2013rza}.\footnote{The complexification of $k$ considered in \cite{ Anninos:2013rza} is not realized in our construction.} Because the parity-violating Vasiliev theories do not exist in Euclidean signature, our construction necessarily  involves an AdS$_4\to$dS$_4$ continuation, in contrast to the EAdS$_4 \to$dS$_4$ continuation employed in \cite{Anninos:2011ui}.

\begin{table}[!h]
\caption{\bf{CONJECTURED dS$_{4}$/CFT$_{3}$ DUALITIES} }
\centering
\begin{tabular}{p{6cm} p{5cm} p{3cm} }
\hline
\bf{VASILIEV dS$_4$ THEORY} & \bf{BOUNDARY CFT$_3$} & \bf{SPECTRUM}  \\ 
\hline\hline
Non-minimal; $\theta_{0}=\frac{\pi}{2}\lambda$
 & $U(N)_k$ Chern-Simons; & \multirow{2}{*}{All integer spins} \\
Neumann scalar & free anticommuting scalar\\ \hline
Non-minimal; $\theta_{0}=\frac{\pi}{2}(1-\lambda)$ & $U(N)_k$ Chern-Simons;  & \multirow{2}{*}{All integer spins} \\ Dirichlet scalar & free commuting  spinor \\ \hline
Non-Minimal; $\theta_{0}=0$ &  $U(N)_\infty$ Chern-Simons & \multirow{2}{*}{All integer spins} \\ Neumann scalar & free anticommuting scalar \\ \hline
Non-minimal; $\theta_{0}={\pi \over 2}$ &  $U(N)_\infty$ Chern-Simons; & \multirow{2}{*}{All integer spins} \\ Dirichlet scalar & free commuting spinor \\ \hline
Minimal; $\theta_{0}=\frac{\pi}{2}\lambda$ & $Sp(N)_k$ Chern-Simons; & \multirow{2}{*}{Even spins} \\ Neumann scalar & free anticommuting scalar \\ \hline
 Minimal; $\theta_{0}=\frac{\pi}{2}(1-\lambda)$ & $Sp(N)_k$ Chern-Simons; & \multirow{2}{*}{Even spins} \\ Dirichlet scalar & free commuting spinor \\ \hline
Minimal $\theta_{0}=0$ & $Sp(N)_\infty$ Chern-Simons; & \multirow{2}{*}{Even spins} \\ Neumann scalar & free anticommuting scalar  \\ \hline
 Minimal; $\theta_{0}=\frac{\pi}{2}$ & $Sp(N)_\infty$ Chern-Simons;& \multirow{2}{*}{Even spins} \\ Dirichlet scalar &  free commuting spinor \\ \hline
\end{tabular}
{$g_{dS}^2={1 \over N}$; $\lambda={N \over k}$; Dirichlet (Neuman) bulk  scalars have  $\Delta=2$ ($\Delta=1$).~~~~~~~~~~~~~~~~~~\\
 Free (critical) anticommuting scalars are dual to critical (free) commuting spinors. }

\end{table}

The outline of this paper is as follows. In Section 2, we review the  Chern-Simons theories coupled to bosonic scalar or fermionic spinor matter in the fundamental of $U(N)$. We also discuss the statistics-reversed versions of these theories, namely Chern-Simons coupled to fundamental anticommuting scalar or commuting spinor matter, as well as Wick rotation from Minkowski to Euclidean space. In Section 3, we review the  parity-violating Vasiliev theories in the AdS$_4$ and dS$_4$ vacua. %In section 4, we discuss the unitarity of Vasiliev theory in dS$_4$ vacuum. 
In section 4, we present an analytic continuation that relates them. In particular, we show how the $n$-point correlation functions in AdS$_4$ and in dS$_4$ are related by this analytic continuation. In section 5, higher-spin bulk duals are conjectured for the various wrong-statistics Chern-Simons-matter theories. Formulae are given relating the bulk coupling constants and boundary conditions to the boundary level, gauge group and interactions.   Spinor conventions are in the appendix.

\section{$U(N)$ Chern-Simons  theories}
In this section, we briefly review the Chern-Simons scalar and Chern-Simons fermion theories in three dimensions, which are conjectured to be dual to the parity-violating Vasiliev theories in AdS$_4$ \cite{Giombi:2011kc,Aharony:2011jz}. We also discuss the statistics reversed versions of these theories which are conjectured, in Section 4, to be dual to the parity-violating Vasiliev theories in dS$_4$.

\subsection{Spinors}
The action for  a complex anticommuting fermion $\psi$ in the fundamental representation of $U(N)$ coupled to a gauge field $A_i$ with a level $k$ Chern-Simons interaction in three Lorentzian dimensions is
\ie\label{cm}
S={k\over 4\pi}\int \tr\left(A dA+{2\over 3}A^3\right)+\int d^3x~\bar\psi\gamma^i D_i\psi.
\fe
According to \cite{Giombi:2011kc}, the spectrum of primary operators in this theory consists of a spin-$s$ single trace operator $J^{(s)}$ for each $s>0$, which take the schematic form
\ie
J^{(s)}_{i_1\cdots i_s}=i^s\bar \psi \gamma_{i_1}D_{i_2}\cdots D_{i_s}\psi+\cdots.
\fe
$J^{(s)}_{i_1\cdots i_s}$ are almost conserved, in the sense that the violation of current conservation is suppressed by a power of $1/N$, and the conformal dimensions of these operators are given by the unitarity bound up to $1/N$ corrections, i.e. $\Delta=s+1+\cO(1/N)$.  In additional to the spin-$s$ primary operators, there is a spin zero primary operator:
\ie 
J^{(0)}=\bar\psi\psi,
\fe
with conformal dimension $\Delta=2+\cO(1/N)$. All other primaries are products of these ``single-trace" operators. This theory is conjectured by \cite{Giombi:2011kc} to be dual to a Vasiliev theory with parity-violating phase $\theta_0$.\footnote{We review parity-violating Vasiliev theories in section 3.} The higher-spin currents $J^{(s)}$ are dual to the higher-spin gauge fields in the bulk, and the spin zero operator $J^{(0)}$ is dual to a bulk scalar with $\Delta=2$ (Dirichlet) boundary condition. The 't Hooft coupling 
\ie \lambda={ N \over k} \fe is mapped by the duality to the parity-violating phase $\theta_0$ by \ie \theta_0={\pi\over 2}(1-\lambda).\fe The planar three-point functions for the spin-$0,1$ currents have been computed in \cite{GurAri:2012is}, which exactly match with the tree level correlation functions in the bulk Vasiliev theory. In general, the $N$ dependence of a Feynman diagram is given by $N^{2-2g-h}$, where $h$ is the number of fermion loops, and $g$ is the genus of the diagram. More explicitly, the $n$-point function takes the form as $N^{2-2g-h}f_{g,h}(\lambda,x^i_k)$, where $x^i_k$ for $k=1,\cdots,n$ are the positions of the $n$ points.

Now consider the same theory, but with opposite statistics for the spinors. The action for the theory is
\ie
S={k\over 4\pi}\int \tr\left(A dA+{2\over 3}A^3\right)+\int d^3x~\bar\xi\gamma^i D_i\xi,
\fe
where $\xi$ is a $commuting$ Dirac spinor in the fundamental representation of $U(N)$. In the 't Hooft large-$N$ limit, the spectrum of single-trace primary operators contains the spin-$s$ operators $J^{(s)}$ for each $s\ge0$. These take the schematic form:
\ie
J^{(s)}_{i_1\cdots i_s}=i^s\bar \xi \gamma_{i_1}D_{i_2}\cdots D_{i_s}\xi+\cdots,\text{  for $s>0$, and  }J^{(0)}=\bar\xi\xi.
\fe
By the same argument as in \cite{Giombi:2011kc}, these spin-$s$ operators are almost conserved and have an anomalous dimension of order $1/N$. The correlation functions of these operators can be computed by the exact same diagrams as in the theory (\ref{cm}) with the anticommuting spinors. The only change is that there is an extra minus sign associated with every independent matter loop by Bose statistics. As a result, the correlation functions with $h$ matter loops at genus $g$ take the form $(-1)^hN^{2-2g-h}f_{g,h}(\lambda,x^i_k)$, where $f_{g,h}(\lambda,x^i_k)$ is the same function as in the theory with anticommuting spinors. So to obtain the current correlation functions in the reversed statistics theory, we simply have to flip the sign of $N$, while keeping $\lambda$ fixed.

\subsection{Scalars}
The Lorentzian action for a three-dimensional complex scalar $\phi$ in the fundamental representation of $U(N)$ coupled to a gauge field $A_i$ with a Chern-Simons interaction at level $k$ is \ie
S={k\over 4\pi}\int \tr\left(A dA+{2\over 3}A^3\right)+\int d^3x\left(|D_i\phi|^2+{\lambda_6\over 3!N^2}(\phi^\dagger\phi)^3\right),
\fe
where $D_i=\partial_i+A_i$, and $k\in\bZ$. %In the t' Hooft large $N$ limit, keeping $\lambda={N\over k}$ and $\lambda_6$ fixed, we could rescale the scalar field $\phi$ as $\phi\to\sqrt{N}\phi$, and the action becomes
%\ie
%S=N\left[{1\over 4\pi\lambda}\int \tr\left(A dA+{2\over 3}A^3\right)+\int d^3x\left(|D_\m\phi|^2+{\lambda_6\over 3!}(\phi^\dagger\phi)^3\right)\right].
%\fe
We are interested in the 't Hooft  large-$N$ limit, keeping $\lambda={N\over k}$ and $\lambda_6$ fixed. According to \cite{Aharony:2011jz}, conformality constrains the parameter $\lambda_6$ to be a function of $\lambda$. The spectrum of operators in the theory includes a single primary operator for each integer spin $s\ge0$. Each $J^{(s)}$ can be written as a symmetric, traceless tensor that is schematically given by
\ie
J^{(s)}_{i_1\cdots i_s}=i^s\phi^\dagger D_{i_1}\cdots D_{i_s}\phi+\cdots.
\fe
As in the fermion case, the $J^{(s)}_{i_1\cdots i_s}$ are almost conserved currents. This theory is conjectured \cite{Aharony:2011jz} to be dual to a parity-violating Vasiliev theory with $\Delta=1$ (Neumann)  boundary condition for the bulk scalar, and the higher-spin operators $J^{(s)}$ are dual to the higher-spin gauge fields in the bulk. The planar three-point functions for spin-$0,1,2$ currents have been computed in \cite{Aharony:2012nh}, which exactly match with the tree level correlation functions in the bulk Vasiliev theory with $\theta_0={\pi\over 2}\lambda$. As in the previous subsection, we want to have a formula for the $N$ dependence of general Feynman diagrams. For this purpose, it is convenient to introduce the auxiliary fields $D$ and $\sigma$. The equivalent action with the auxiliary fields is given by
\ie
S={k\over 4\pi}\int \tr\left(A dA+{2\over 3}A^3\right)+\int d^3x\left(|D_i\phi|^2+\phi^\dagger(\sigma^2-D)\phi+\sqrt{3!N^2\over \lambda_6}\tr( D\sigma)\right).
\fe
In this form it is evident that the $N$ dependence of the Feynman diagrams with $h$ matter loops at genus $g$ is given by $N^{2-2g-h}$.

Now consider the same theory, but with opposite statistics for the scalar field. The action for the theory is
\ie
S={k\over 4\pi}\int \tr\left(A dA+{2\over 3}A^3\right)+\int d^3x\left(|D_i\chi|^2+{\lambda_6\over 3!N^2}(\chi^\dagger\chi)^3\right),
\fe
where $\chi$ is an anticommuting scalar in the fundamental representation of $U(N)$. In the 't Hooft large-$N$ limit, the spectrum of single-trace primary operators contains the spin-$s$ operators $J^{(s)}$ for each $s\ge0$. These take the schematic form:
\ie
J^{(s)}_{i_1\cdots i_s}=i^s\chi^\dagger D_{ i_1}\cdots D_{ i_s}\chi+\cdots.
\fe
By the same argument as in \cite{Aharony:2011jz}, these spin-$s$ operators are almost conserved and have an anomalous dimension of order $1/N$. The correlation functions of these operators can be computed by the exact same diagrams as in the theory with the commuting scalar. The only change is that there is an extra minus sign associated with every matter loop by Fermi statistics. The net effect is to flip the sign of $N$ while keeping $\lambda$ fixed.

\subsection{Wick rotation}
The future boundary of dS$_4$ has Euclidean signature, so we are interested in Euclidean CFT$_3$s.  Let us consider the analytic continuation of the statistics reversed $U(N)$ Chern-Simons spinor and Chern-Simons scalar theories from Lorentzian signature to Euclidean signature. This can easily be done by an analytic continuation of the coordinate of the function $f_{g,h}(\lambda,x^i_a)$ from $x^0_a$ to $-ix^3_a$. More explicitly, the analytic continuation of the higher-spin currents are
\ie \label{dg}
J^{(s),E}_{i_1\cdots i_s}=i^n J^{(s),L}_{i_1\cdots i_s}\Big|_{x^0\to -ix^3},
\fe
where the superscripts $L,E$ distinguish the operators in Lorentzian or Euclidean signature, and $n$ is the number of the indices of $J^{(s),L}_{i_1\cdots i_s}$ that are $0$. It is convenient to define generating functions
\ie
&J^{(s)}_L(x|y)=\sum_{i_1,\cdots,i_s}J^{(s),L}_{i_1\cdots i_s}(y\sigma_L^{i_1}y)\cdots(y\sigma_L^{i_s}y),
\\
&J^{(s)}_E(x|y)=\sum_{i_1,\cdots,i_s}J^{(s),E}_{i_1\cdots i_s}(y\sigma_E^{i_1}y)\cdots(y\sigma_E^{i_s}y),
\fe
where $y^\A$ is an auxiliary bosonic spinor variable, $y\sigma^{i}_Ly=y_\A(\sigma^{i}_L)^\A{}_{\B}y^\B$, $y\sigma_E^{i}y=y_\A(\sigma^{i}_L)^\A{}_{\B}y^\B$, $(\sigma^{i}_E)^\A{}_{\B}=(\sigma^1,\sigma^3,\sigma^2)$ are Pauli matrices, and $(\sigma^{i}_L)^\A{}_{\B}=(\sigma^1,\sigma^3,i\sigma^2)$. In terms of the generating functions, the analytic continuation of the higher-spin currents can be simply stated as
\ie
J^{(s)}_E(x|y)=J^{(s)}_L(x|y)\Big|_{x^0\to -ix^3},
\fe
which accounts for the $i^n$ in (\ref{dg}). 
The analytic continuation of the correlators is simply 
\ie
\vev{J^{(s_1)}_E(x_1|y)\cdots J^{(s_n)}_E(x_n|y)}=\vev{J^{(s_1)}_L(x_1|y)\cdots J^{(s_n)}_L(x_n|y)}\Big|_{x^0_k\to -ix^3_k}.
\fe

\section{Vasiliev theories in AdS$_4$ and dS$_4$}
In this section, we review the Vasiliev theory in AdS$_4$ and dS$_4$ backgrounds \cite{Vasiliev:1992av,Vasiliev:1995dn,Vasiliev:1999ba,Iazeolla:2007wt,Anninos:2011ui}. We will start with a background independent formalism, and then specify the vacuum solutions and reality conditions. The fields in the Vasiliev theory are functions of bosonic variables $(x,Y,Z)=(x^\m,y^\A,z^\A,\bar y^{\dot \A},\bar z^{\dot \A})$. Here $x^\mu$  are an arbitrary set of coordinates on the four dimensional spacetime manifold with signature $(-,+,+,+)$. $(y^\A,z^\A,\bar y^{\dot \A},\bar z^{\dot \A})$ are commuting $SO(1,3)$ spinors.
%As discussed in \cite{Iazeolla:2007} we can expand Vasiliev's equations about vacuum dS$_4$ and $EAdS_4$ solutions. The reality properties of the spinors depend upon the signature of the space time. We refer the reader to \cite{Iazeolla:2007} for a listing of these reality properties in different signatures and for more details on Vasiliev's theory. 
Our spinor conventions for AdS$_4$ and dS$_4$ are given in the Appendix A. The Vasiliev master fields consist of an $x$-space 1-form
\ie
W=W_\m dx^\m,
\fe
a $Z$-space 1-form
\ie
S=S_\A dz^\A+S_{\dot \A} d\bar z^{\dot \A},
\fe
and a scalar $B$, all of which depend on all the bosonic variables introduced above. The master fields are  truncated by the condition
\ie\label{trunc}
\pi\bar\pi(W)=W,~~~\pi\bar\pi( S)=S,~~~\pi\bar\pi( B)=B,
\fe
where the $\pi$-action is defined as $\pi:(y,z,dz,\bar y,\bar z,d\bar z)\mapsto(-y,-z,-dz,\bar y,\bar z,d\bar z)$, and the $\bar \pi$-action is given by the $\pi$-action with exchanging the barred and unbarred variables. It is easy to check that the equation of motion is consistent with the truncation (\ref{trunc}).

The master field equation of motion is \cite{Iazeolla:2007wt,Chang:2012kt}:
\ie\label{eom}
d_x \hat\cA+\hat\cA*\hat\cA=\left({1\over 4}+B*K~e^{i\theta_0}\right)dz^2+ \left({1\over 4}+B*\overline{K}~e^{-i\theta_0}\right)d\bar z^2,
\fe
where $\hat\cA=W+S-{1\over 2}zdz$ and $K=e^{zy},\overline K=e^{\bar z\bar y}$ and $\theta_0$ is a coupling constant and $d_x$ is the exterior derivative with respect to spacetime coordinate $x^\m$. Here the Vasiliev's $*$-product is defined by
\ie
f*g
=f(Y,Z) exp\left[\epsilon^{\alpha\beta}\left(\overleftarrow{\partial}_{y^{\alpha}}+\overleftarrow{\partial}_{z^{\alpha}}\right)\left(\overrightarrow{\partial}_{y^{\beta}}-\overrightarrow{\partial}_{z^{\beta}}\right)+\epsilon^{\dot{\alpha}\dot{\beta}}\left(\overleftarrow{\partial}_{y^{\dot{\alpha}}}+\overleftarrow{\partial}_{z^{\dot{\alpha}}}\right)\left(\overrightarrow{\partial}_{y^{\dot{\beta}}}-\overrightarrow{\partial}_{z^{\dot{\beta}}}\right)\right]g(Y,Z).
\fe

In the parity invariant A-type and B-type theories, $\theta_0$ takes the values $\theta_0=0$ and $\theta_0={\pi\over 2}$, respectively. Parity is not conserved for generic $\theta_0$. In addition to $\theta_0$, the quantum Vasiliev theory has an additional coupling constant $g$ which measures the strength of quantum corrections. For the Vasiliev theory on AdS$_4$ and dS$_4$ background, we will denote this coupling as $g_{AdS}$ or $g_{dS}$, respectively.

The Vasiliev master fields are, a priori, complex-valued fields. There are several different consistent reality conditions that can be imposed on the master fields. Different reality conditions preserve different vacuum solutions. In the following two subsections, we review the Vasiliev theory on AdS$_4$ and dS$_4$ backgrounds, and specify the reality conditions that preserve these two backgrounds. 

\subsection{AdS$_4$ }

Let us consider the Vasiliev theory with the spacetime signature $(+,+,+,-)$, with coordinates denoted  $x^\m=(z,x^1,x^2,x^0)$. The AdS$_4$ vacuum solution is 
\ie
W=W_0=\omega_0(x|Y)+e_0(x|Y),~~~~B=0,~~~~S=0,
\fe
where 
\ie\label{AdS}
&\omega_0(x|Y) = -{1\over  8} {dx_i\over z}\left( y\sigma_{AdS}^{iz}y + \bar y\sigma_{AdS}^{iz}\bar y \right),\\
& e_0(x|Y) = -{1\over 4} {dx_\mu\over z} y\sigma_{AdS}^\mu \bar y,
\fe
and the $\sigma$-matrices are defined in the \eqref{SADS}. The metric or vielbein are not fundamental quantities in Vasiliev theory. They can be extracted from the vacuum solution $W_0$. The vielbein $e^a_{AdS}$ can be extracted from $e_0(x|Y)$ by
\ie
e_0(x|Y)=-{1\over 4}\eta_{ab}e_{AdS}^a(y\sigma_{AdS}^b\bar y),
\fe
and the AdS$_4$ metric is then given by $g^{AdS}_{\m\n}dx^\m dx^\n=\eta_{ab}e^a_{AdS} e^b_{AdS}=({\eta_{\m\n} /z^2})dx^\m dx^\n$.

The reality condition on the Vasiliev's master fields that preserves the AdS$_4$ vacuum solution \label{AdS4} is\footnote{The $\iota(W)^*$ is to be understood as first acting the $\iota$ on $W$ then taking the complex conjugate.}
\ie\label{rc}
-\iota(\hat\cA)^*=\hat\cA,~~~\iota \pi ( B)^*=B,
\fe
% or the reality condition on $B$ can be rewritten as
%\ie
%\iota(B)^* = \overline K*B* \overline K\Gamma,
%\fe
where the reality condition on the auxiliary variables $(Y,Z)$ are defined in appendix A, and the $\iota$-action is defined as $\iota:(y,\bar y,z,\bar z,dz,d\bar z)\mapsto(iy,i\bar y,-iz, -i\bar z,-i dz,-i d\bar z)$. It follows that the $\iota$-action would reverse the $*$-product, i.e.
\ie
\iota(f(Y,Z)*g(Y,Z))=\iota(g(Y,Z))*\iota(f(Y,Z)).
\fe

At the linear level, after an appropriate gauge fixing and eliminations of the auxiliary fields, the Vasiliev's equation of motion on the background \eqref{AdS4} reduces to the Fronsdal's equation of motion \cite{Vasiliev:2003ev,Bekaert:2005vh,Giombi:2009wh,Giombi:2012ms}. The Fronsdal equation in the traceless gauge is, 
\ie\label{linten}
-(\Box-m^2)\varphi^{AdS}_{\mu_1\cdots\mu_s} + s\nabla_{(\mu_1}\nabla^\nu \varphi^{AdS}_{\mu_2\cdots\mu_s)\nu}
-{s(s-1)\over 2(d+2s-3)} g^{AdS}_{(\mu_1\mu_2}\nabla^{\nu_1}\nabla^{\nu_2}\varphi^{AdS}_{\mu_3\cdots\mu_s)\nu_1\nu_2} = 0,
\fe
%\ie\label{linten}
%&(\Box-m^2)\varphi^{AdS}_{\m_1\cdots\m_s}-s \p_{({\m_1}}\p^\n\varphi^{AdS}_{{\m_2\cdots\m_s})\n}+{1\over 2}s(s-1)\p_{(\m_1}\p_{\m_2}\varphi^{AdS}_{{\m_3\cdots\m_s})\n}{}^{\n}
%-s(s-1)g^{AdS}_{({\m_1\m_2}}\varphi^{AdS}_{{\m_3\cdots\m_s})\n}{}^{\n}=0,
%\fe
where $m^2=s(s-2)-2$, and $\varphi^{AdS}_{\m_1\cdots\m_s}$ is traceless symmetric spin-$s$ gauge field. It appears in the components of the Vasiliev master fields $W,B$. More explicitly, the spin-$s$ higher-spin gauge field $\varphi^{AdS}_{\m_1 \m_2\cdots \m_s}$ is an expansion coefficient of the master field $W$  (equation (3.59) in \cite{Giombi:2009wh})
\ie\label{WtoO}
W(x,Y,Z=0)\big|_{y^{s-1},\bar y^{s-1}}
%\label{OS1S1}
%&\propto z^{s-1}(\sigma_{AdS,\m})^{\A\dot\B}\partial_\A\partial_{\dot\B}\Phi^s_{AdS}(x|y,\bar y)
%\\
%&={(s!)^2\over 2\tilde N_s(2s)!}\sum e^{a_1}_{\m}\varphi_{a_1 a_2\cdots a_s}(x,z)Y^{a_2}\cdots Y^{a_s}
%\\
&\propto (iz)^{s-1} \varphi^{AdS}_{\m_1 \m_2\cdots \m_s}(y\sigma_{AdS}^{\m_2}\bar y)\cdots (y\sigma_{AdS}^{\m_s}\bar y).
\fe
The spin zero field $\varphi_{AdS}$ sits in the $(Y,Z)$ independent part of master field $B$ as
\ie\label{BVP}
B(x|Y,Z)\Big|_{Y=Z=0}=\varphi_{AdS}.
\fe
At the nonlinear level, one can, in principle, extract the corrections to the right hand side of the linear equation \eqref{linten} from the Vasiliev equation \label{eom} order by order in the number of higher spin gauge fields. A systematic procedure for this was discussed in \cite{Sezgin:2002ru,Giombi:2009wh}. %It is widely expected that this nonlinear second order equation of motion can be derived from the variation of an action.

The reality condition \eqref{rc} for the master fields, hence, gives the reality condition on the physical higher-spin gauge fields. More explicitly, by imposing the reality condition on equation \eqref{WtoO}, we find
\ie\label{RCPF}
(\varphi^{AdS}_{\m_1\cdots\m_s})^*=\varphi^{AdS}_{\m_1\cdots\m_s}.
\fe
The scalar field, on the other hand, is the bottom component of  $B$ according to \eqref{BVP}. Imposing the reality condition on \eqref{BVP} gives $\varphi^*_{AdS}=\varphi_{AdS}$ for the spin-0 field.

The scalar has mass square $m^2=-2$. Depending on the boundary condition for this scalar, its dual operator has either dimension $\Delta= 1$ or $\Delta = 2$, classically. We will refer to the two different boundary conditions as $\Delta= 1$ (Neuman) and $\Delta= 2$ (Dirichlet) boundary conditions, respectively.

\subsection{dS$_4$ }
In dS$_4$, we label the coordinates by $(\eta,x^1,x^2,x^3)$ with the signature $(-,+,+,+)$. The dS$_4$ vacuum solution to Vasiliev's equation of motion is given by
\ie
W=W_0=\omega_0(x|Y)+e_0(x|Y),~~~~B=0,~~~~S=0,
\fe
and
\ie\label{dS}
&\omega_0(x|Y) =- {1\over  8} {dx_i\over \eta}\left( y \sigma_{dS}^{i\eta}y + \bar y\sigma_{dS}^{i\eta}\bar y \right),\\
& e_0(x|Y) = -{i\over 4} {dx_\mu\over \eta} y\sigma_{dS}^\mu \bar y,
\fe
where the $\sigma_{dS}^\mu,\sigma_{dS}^{i\eta}$ are the $\sigma$-matrices defined in (\ref{dssig}). The vielbein $e^a_{dS}$ is extracted from $e_0(x|Y)$ according to
\ie
e_0(x|Y)=-{1\over 4}\eta^{dS}_{ab}e_{dS}^a(y \sigma_{dS}^b\bar y).
\fe
The metric is then $g^{dS}_{\m\n}dx^\m dx^\n=\eta_{ab}e_{dS}^ae_{dS}^b=-(\eta_{\m\n}/ \eta^2)dx^\m dx^\n$.
A reality condition on the Vasiliev's master fields that preserves the dS$_4$ vacuum solution \eqref{dS} is
\ie\label{RCDS1}
\pi(\hat\cA)^*=\hat\cA,~~~\pi(B)^*=B,
\fe
%and
%\ie\label{RCDS2}
%-\iota\pi(\hat\cA)^*=\hat\cA,~~~\iota(B)^*=B.
%\fe
which is also compatible with the equation of motion \eqref{eom} and truncation \eqref{trunc}.\footnote{This reality condition agrees with \cite{Iazeolla:2007wt} when reduced to the minimal theory.}

As in the AdS$_4$ case, the linearized Vasiliev equation of motion on the dS$_4$ background \eqref{dS} is reduced to the Fronsdal equation \eqref{linten} with all the subscripts and superscripts $AdS$ replaced by $dS$. The spin-$s$ higher-spin gauge field $\varphi^{dS}_{\m_1\cdots\m_s}$ is the expansion coefficient of the master fields $W$ and $B$:
\ie
&W(x,Y,Z=0)\big|_{y^{s-1},\bar y^{s-1}}
\propto \eta^{s-1} \varphi^{dS}_{\m_1 \m_2\cdots \m_s}(y\sigma_{dS}^{\m_2}\bar y)\cdots (y\sigma_{dS}^{\m_s}\bar y),
\\
&B(x|Y,Z)\Big|_{Y=Z=0}=\varphi_{dS}.
\fe
The reality condition \eqref{RCDS1} implies
\ie\label{HSRC1}
(\varphi^{dS}_{\m_1\cdots\m_s})^*=(-1)^s\varphi^{dS}_{\m_1\cdots\m_s}.
\fe Note that, for the odd spin gauge fields, this reality condition differs from the reality condition \eqref{RCPF} in AdS$_4$. However we will find below that they are mapped into one another by our analytic continuation procedure.

\section{Analytic continuation from AdS$_4$ to dS$_4$}
In this section, we describe the analytic continuation of Vasiliev theory from AdS$_4$ to dS$_4$.\footnote{We could have obtained the dualities for the parity invariant Type A and B models by analytically continuing the corresponding results from $EAdS_4$ instead of AdS$_4$. This was done for the minimal Type A model in \cite{Anninos:2011ui}. But because of the $(4,0)$ signature of EAdS$_4$ we cannot impose reality conditions on the auxiliary Weyl spinors $(y,\bar y, z, \bar z)$. As noted in \cite{Iazeolla:2007wt} this means that the reality conditions on the master fields in $EAdS_4$ are not compatible with Vasiliev's equation for the parity-violating models. Here we have circumvented this Euclidean problem by directly continuing from Lorentzian AdS$_4$ to Lorentzian dS$_4$.} Let us start with the Vasiliev equation - with any value of $\theta_0$ - for the master fields expanded about the  AdS$_4$ background. Before imposing reality conditions on either the auxilliary spinor variables or the master fields, the analytic continuation of the coordinates:
\ie\label{ACco}
( z, x^1,x^2,x^0)_{AdS}=(-i\eta,x^1,x^2,-ix^3)_{dS},
\fe
maps the AdS$_4$ background solution \eqref{AdS} to the dS$_4$ background solution \eqref{dS}. This gives the first-order Vasiliev master field equations expanded about the dS$_4$ background. It follows that the second-order equation of motion for the physical higher-spin component fields in dS$_4$ obtained by continuing the second-order Vasiliev equation in AdS$_4$ will match the second-order equation obtained directly from the Vasiliev equation expanded about the dS$_4$ background. Hence the AdS$_4$ and dS$_4$ theories are simply related by analytic continuation. We will describe our prescription for the analytic continuation of the higher-spin gauge fields in subsection \ref{Achsf}, and of the  correlation functions in subsection \ref{Accf}.

\subsection{Fields}\label{Achsf}
In this subsection, we give the analytic continuation of higher-spin gauge fields. First of all, applying the analytic continuation \eqref{ACco} to the background solutions \eqref{AdS} and \eqref{dS}, we find that the AdS$_4$ metric $g^{AdS}_{\m\n}$ is indeed related to the dS$_4$ metric $g^{dS}_{\m\n}$ by
\ie
g^{dS}_{\m\n}dx^\m dx^\n=g^{AdS}_{\m\n}dx^\m dx^\n\Big|_{z\to-i\eta}^{x^0\to-ix^3}.
\fe
The prescription for continuing the higher-spin fields  is
\ie\label{ACPCDVAR}
\varphi^{dS}_{\m_1\cdots\m_s}(\eta,x^1,x^2,x^3)&=i^{n}\varphi^{AdS}_{\m_1\cdots\m_s}( -i\eta,x^1,x^2,- ix^3),
\fe
where $n$ is the total number of $0$ and $z$ indices. By the reality conditions \eqref{RCPF} and \eqref{HSRC1}, the odd spin fields are pure imaginary on the left hand side, while the odd spin fields are real on the right hand side. 

At first sight, this analytic continuation might seem to  lead to bunch of unwanted $i$'s in the Vasiliev equation of motion, but this is actually not the case. Note that there are no explicit indices in Vasiliev's equation of motion, that is, every free index on a higher-spin gauge field must be contracted with $y\sigma_{AdS}^\m\bar y$, $y\sigma_{AdS}^{\m\n}y$, $\bar y\sigma_{AdS}^{\m\n}\bar y$ or similar terms with $y,\bar y$ replaced by $z,\bar z$. When we perform the analytic continuation from AdS$_4$ to dS$_4$, the $\sigma$-matrices for AdS$_4$  absorb the $i$'s and turn into the $\sigma$-matrices for dS$_4$. More explicitly, the first and last components of $\sigma_{AdS}^\m$ in $\sigma^\m_{AdS,\A\dot \B}=(i{\bf 1},\sigma^1,\sigma^3,i\sigma^2)$ absorb an $-i$ and turn into the first and last components of $\sigma_{dS}^\m$ in $\sigma^\m_{dS,\A\dot \B}=({\bf 1},\sigma^1,\sigma^3,\sigma^2)$. This suggests that we focus on the generating functions:
\ie\label{GTF}
\Phi^s_{AdS}(x|y,\bar y)=\varphi^{AdS}_{\m_1\cdots \m_s}(y\sigma_{AdS}^{\m_1}\bar y)\cdots (y\sigma_{AdS}^{\m_s}\bar y),
\fe
and 
\ie
\Phi^s_{dS}(x|y,\bar y)=\varphi^{dS}_{\m_1\cdots \m_s}(y\sigma_{dS}^{\m_1}\bar y)\cdots (y\sigma_{dS}^{\m_s}\bar y).
\fe
For these the analytic continuation procedure takes the very simple form
\ie\label{ACPHIDA}
\Phi^s_{dS}(\eta,x^1,x^2,x^3|Y)&=\Phi^s_{AdS}( -i\eta,x^1,x^2,- ix^3|Y).
\fe

%The spacetime signature is $(-,+,+,+)$.  The reality condition on the Vasiliev's master fields $W$ and $S$ are defined as\cite{Iazeolla:2007}
%\ie\label{rc}
%\iota(W)^*=-W,~~~\iota(S)^* = -S
%\fe
%The consistent reality condition on $B$ differs by a sign for the Type A and B models in dS$_4$. For the Type A case we have, $\iota \pi(B)^* = B$ while for Type B it is $\iota \pi (B)^* =-B$\footnote{The $\iota(W)^*$ is to be understood as first acting the $\iota$ on $W$ then taking the complex conjugate.}

%Here the $\iota$-action is defined as $\iota:(y,\bar y,z,\bar z,dz,d\bar z)\mapsto(iy,i\bar y,-iz, -i\bar z,-i dz,-i d\bar z)$. It follows that the $\iota$-action reverses the $*$-product \cite{Chang:2012kt}. Noting that the spacetime coordinates $x^\m$ are real under the complex conjugate it is straightforward to check that the Vasiliev equation of motion \eqref{eom} is compatible with this reality condition.

%The dS$_4$ vacuum solution to Vasiliev's equation of motion is given by
%\ie
%W=W_0=\omega_0(x|Y)+e_0(x|Y),~~~~B=0,~~~~S=0,
%\fe
%and
%\ie\label{dS4}
%&\omega_0(x|Y) =- {1\over  8} {dx_i\over \eta}\left( y\sigma^{i\eta}y + \bar y\sigma^{i\eta}\bar y \right),\\
%& e_0(x|Y) = -{1\over 4} {l_{dS}dx_\mu\over \eta} y\sigma^\mu \bar y,
%\fe
%where the $\sigma$-matrices and their reality properties are defined in the Appendix.
% (\ref{dssig}). 

\subsection{Correlators}\label{Accf}

Now we give the prescription for analytically continuing the AdS$_4$ correlators to the dS$_4$ ones. In order to go  from the classical equations of motion to the quantum ones one must specify an additional coupling (essentially $\hbar$), which we have denoted $g_{AdS}$ ($g_{dS}$) for AdS$_4$ (dS$_4$).  These couplings may be defined as the coefficient of the singularity in the scalar two point function:  More explicitly, one needs to associate a factor $g_{AdS}^{-2}$ with each internal or external line, and a factor $g^{2}_{AdS}$ with each (cubic) vertex. This gives a factor $g_{AdS}^{2n+2\ell-2}$ for the $\ell$-loop, $n$-point function. For example, the bulk scalar two point function takes the form 
\ie\label{adsc}
\vev{\varphi^0_{AdS}(x_1^\m)\varphi^0_{AdS}(x_2^\m)}_{AdS}\approx g_{AdS}^2{z_1 z_2\over -(x^0_1-x^0_2)^2 +(x^1_1-x^1_2)^2+(x^2_1-x^2_2)^2+(z_1-z_2)^2},
\fe
\ie\label{dsc}
\vev{\varphi^0_{dS}(x_1^\m)\varphi^0_{dS}(x_2^\m)}_{dS}\approx g_{dS}^2{\eta_1 \eta_2\over -(\eta_1-\eta_2)^2
+  (x^1_1-x^1_2)^2+(x^2_1-x^2_2)^2+(x^3_1-x^3_2)^2},
\fe
in the limit when the two points are very close to each other, i.e. $(z_1,\vec{x}_1)\to(z_2,\vec{x}_2)$. Once this normalization is specified, the dependence of higher point correlators on the coupling is determined by unitarity. By our analytic continuation procedure, the $z_1z_2$ in the numerator of (\ref{adsc}) becomes $-\eta_1\eta_2$ in the numerator of (\ref{dsc}).  Hence, as in \cite{Anninos:2011ui}, the bulk coupling constant must continue as $g^2_{AdS}\to -g^2_{dS}$ at the same time to maintain the positivity of the kinetic term. 

 We now examine the short distance singularity in the two point function for fields of higher spin gauge fields. For $s>0$, the two point functions for the physical transverse components of two higher spin gauge fields have the singularity
\ie\label{adsc}
&\vev{\varphi^{AdS}_{i_{1}\cdots i_{s}}(x_1^\m)\varphi^{AdS}_{i_{1}\cdots i_{s}}(x_2^\m)}_{AdS}\approx g_{AdS}^2{(z_{1}z_{2})^{-s+1}\over -(x^0_1-x^0_2)^2 +(x^1_1-x^1_2)^2+(x^2_1-x^2_2)^2+(z_1-z_2)^2},
\\
&\vev{\varphi^{dS}_{i_{1}\cdots i_{s}}(x_1^\m)\varphi^{dS}_{i_{1}\cdots i_{s}}(x_2^\m)}_{dS}\approx g_{dS}^2{(-1)^s(\eta_{1}\eta_{2})^{-s+1}\over-(\eta_1-\eta_2)^2
+  (x^1_1-x^1_2)^2+(x^2_1-x^2_2)^2+(x^3_1-x^3_2)^2},
\fe
in the limit when the two points are very close to each other, where $i_1,i_2,\cdots,i_s=1,2$. They are related by our analytic continuation procedure and the analytic continuation of coupling constant: $g^2_{AdS}\to -g^2_{dS}$. Recalling  that the reality condition implies that the odd spin component fields are purely imaginary in dS$_4$,  the important factor of $(-)^s$ in the second line of (\ref{adsc}) implies positivity of the kinetic term (in terms of real fields) is maintained by the analytic continuation. 

The rule for the analytic continuation of the bulk correlation function is
\ie
\vev{\Phi^{s_1}_{dS}(x_1^\m|Y)\cdots \Phi^{s_n}_{dS}(x_n^\m|Y)}_{dS}=\vev{\Phi^{s_1}_{AdS}(x_1^\m|Y)\cdots \Phi^{s_n}_{AdS}(x_n^\m|Y)}_{AdS}\Big|_{x^0\to -ix^3,z \to -i \eta}^{g^2_{AdS}\to -g^2_{dS}}.
\fe
The boundary correlation functions can be extracted from the bulk correlation functions by taking the scaled boundary limit \cite{Giombi:2009wh,Giombi:2012ms}\footnote{Notice that $-iy\sigma^\m_{AdS}\sigma^z_{AdS} y=\delta_{\m,i}y\sigma^i_Ly$, and $y\sigma^\m_{dS}\sigma^\eta_{dS} y=\delta_{\m,i}y\sigma^i_Ey$.}:
\ie
J^{(s)}_{AdS}(\vec{x}|y)=\lim_{z\to 0}{1\over g_{AdS}^2 z}\Phi^s_{AdS}(x^\m|y,\bar y=-i\sigma^z_{AdS} y),
\fe
and similarly
\ie
J^{(s)}_{dS}(\vec{x}|y)=\lim_{\eta\to 0}{1\over  ig_{dS}^2 \eta}\Phi^s_{dS}(x^\m|y,\bar y=\sigma^\eta_{dS} y).
\fe
Therefore, we have
\ie\label{JCAC}
\vev{J^{(s_1)}_{dS}(\vec{x}_1|y)\cdots J^{(s_n)}_{dS}(\vec{x}_n|y)}_{dS}=\vev{J^{(s_1)}_{AdS}(\vec{x}_1|y)\cdots J^{(s_n)}_{AdS}(\vec{x}_n|y)}_{AdS}\Big|_{x^0\to -ix^3}^{g^2_{AdS}\to -g^2_{dS}}.
\fe
%The $\ell$-loop boundary correlators are proportional to $g^{2\ell-2}_{AdS}$ (or $g^{2\ell-2}_{dS}$ for the theory in dS$_4$). 

\section{dS$_4$/CFT$_3$ }

In section 2, we showed that for both the Chern-Simons scalar and Chern-Simons fermion theories, the net effect of reversing the statistics of the matter fields is flipping the sign of $N$ while keeping $\lambda$ fixed. Correlators of the statistics reversed theories can be further transformed to the corresponding correlators in Euclidean signature by analytic continuation $x^0\to -ix^3$. In section 3, we showed that the correlators in the Vasiliev theory in AdS$_4$ and dS$_4$ are related by the analytic continuation \eqref{JCAC}. In particular, the correlators in dS$_4$ are given by the correlators in AdS$_4$ by flipping the sign of the squared coupling, i.e. $g^2_{AdS}\to -g^2_{dS}$ together with the analytic continuation on the coordinates. Using the conjectures in \cite{Giombi:2011kc,Aharony:2011jz}, the parity-violating Vasiliev theory in AdS$_4$, with the $\Delta=1$ or $\Delta=2$ boundary condition for the scalar, is dual to the Chern-Simons scalar or Chern-Simons fermion theory, respectively, with  $N=g_{AdS}^{-2}$. For the case $\Delta=1$  ($\Delta=2$), the bulk parity-violating phase $\theta_0$ and boundary 'tHooft coupling $\lambda={N\over k}$ are  related by  $\theta_0={\pi\over 2}\lambda$ ($\theta_0={\pi\over 2}(1-\lambda)$). Hence, if the conjectures in \cite{Giombi:2011kc,Aharony:2011jz} are correct, the parity-violating Vasiliev theory in dS$_4$, with either boundary condition, is dual to the statistics reversed Chern-Simons scalar or Chern-Simons spinor theories, respectively, with $N=g^{-2}_{dS}$, with $\theta_0$ and $\lambda$ obeying the same boundary-condition-dependent relation as in the AdS$_4$ theory.

In the special case  $k\to\infty$ of our conjecture, we obtain that the Type A theory in dS$_4$ with $\Delta=1$ boundary condition is dual to the free $U(N)$ anticommuting scalar theory, and the Type B theory in dS$_4$ with $\Delta=2$ boundary condition is dual to the free $U(N)$ commuting spinor theory. 
Our conjecture can also be generalized to the $Sp(N)$ Chern-Simons anticommuting scalar or commuting spinor theories.\footnote{The correlators in the $Sp(2N)$ Chern-Simons theory with wrong-statistics matter are equal to the correlators in the $SO(2N)$ Chern-Simons matter theory with $N$ replaced by $-N$. \cite{Mkrtchian:1981bb}} The bulk dual of these theories is the Vasiliev theory in dS$_4$ background with minimal truncation: $-\iota(\hat \cA)=\hat\cA,\iota\pi(B)=B$.
The Chern-Simons critical scalar and Chern-Simons critical spinor theories are also dual to the parity-violating Vasiliev theory with the $\Delta=2$ or $\Delta=1$ boundary conditions, respectively. On the CFT side, by the bosonization duality \cite{Aharony:2012nh}, the Chern-Simons critical scalar theory is dual to the Chern-Simons non-critical spinor theory, and the Chern-Simons critical spinor theory is dual to the Chern-Simons non-critical scalar theory. We expect the bosonization duality still holds after reversing the statistics of the matter fields.

\section*{Acknowledgements}
We are grateful to Dionysios Anninos, Xi Yin, and especially Gim Seng Ng for useful conversations. This work was supported in part by DOE grant DE-FG02-91ER40654 and the Fundamental Laws Initiative at Harvard.

\appendix
\section{Conventions}
In this appendix, we give our conventions for the $\sigma$-matrices and the auxilliary spinor variables $(y,\bar y,z,\bar z)$ for the theories in dS$_4$ and AdS$_4$.
\subsection{AdS$_4$}
AdS$_4$ has signature $(+,+,+,-)$. It is parametrized by the coordinate $(z,x^1,x^2,x^0)$ in Poincare patch. The $\sigma$-matrices in AdS$_4$ are defined by
\ie\label{sg}
(\sigma^\m_{AdS})_{\A}{}^{\dot \B}(\sigma_{AdS}^\n)^{\gamma}{}_{\dot\B}+(\sigma_{AdS}^\n)_\A{}^{\dot \B}(\sigma_{AdS}^\m)^{\gamma}{}_{\dot\B}=2\delta^\C_\A\eta^{\m\n}_{AdS},
\fe
where $\eta^{\m\n}_{AdS}=\text{diag}(1,1,1,-1)$. An explicit representation\footnote{This is not the conventional representation for the $\sigma$-matrices, but it is related by analytic continuation to the conventional representation for the $\sigma$-matrices in dS$_4$.} of the $\sigma$-matrices is
\ie\label{SADS}
(\sigma_{AdS}^\m)_{\A\dot \B}=(i{\bf 1},\sigma^1,\sigma^3,i\sigma^2),
\fe
where the $\sigma^1,\sigma^2,\sigma^3$ are Pauli matrices. In this representation the complex conjugate of the $\sigma$-matrices are given by
\ie
(\sigma^\m_{AdS,\A\dot \B})^*=-(\sigma_{AdS}^{\m})^{\B\dot\A}.
\fe
The reality condition for the bosonic spinor variables $(Y,Z)$ is defined as 
\ie
(y^\A)^*=\bar y_{\dot \A},~~~~(\bar y_{\dot\A})^*= y^{ \A},~~~~(z^\A)^*=\bar z_{\dot \A},~~~~(\bar z_{\dot\A})^*= z^{ \A},
\fe
such that $y\sigma_{AdS}^\m \bar y$ is real and $(y\sigma_{AdS}^\m\sigma_{AdS}^\n  y)^*=\bar y\sigma_{AdS}^\m\sigma_{AdS}^\n \bar y$. We also define $(\sigma_{AdS}^{\m\n}){}_{\A\B}=(\sigma_{AdS}^{[\underline{\m}})_{\A}{}^{\dot\C}(\sigma_{AdS}^{\underline{\n}]})_{\B\dot\C}$.

\subsection{dS$_4$}
dS$_4$ has signature $(-,+,+,+)$. It is parametrized by the coordinate $(\eta,x^1,x^2,x^3)$ in Poincare patch. Our definition of the $\sigma$-matrices in dS$_4$ is different from the $\sigma$-matrices in AdS$_4$. Hence, we denote the $\sigma$-matrices in dS$_4$ by $\sigma_{dS}$ to avoid confusion. The $\sigma$-matrices in dS$_4$ are defined by the same algebra as in AdS$_4$:
\ie\label{dssig}
(\sigma_{dS}^\m)_\A{}^{\dot \B}(\sigma^\n_{dS})^{\gamma}{}_{\dot\B}+(\sigma_{dS}^\n)_\A{}^{\dot \B}(\sigma^\m_{dS})^{\gamma}{}_{\dot\B}=2\delta^\C_\A\eta^{\m\n}_{dS},
\fe
however, with a different representation:
\ie\label{SDS}
(\sigma_{dS}^\m)_{\A\dot \B}=({\bf 1},\sigma^1,\sigma^3,\sigma^2),
\fe
and $\eta^{\m\n}_{dS}=\text{diag}(-1,1,1,1)$. In this representation, the complex conjugate of the $\sigma$-matrices is given by
\ie
(\sigma^\m_{dS,\A\dot \B})^*=(\sigma_{dS}^\m)_{\B\dot\A}.
\fe
The reality condition for the auxiliary variables is defined as 
\ie
(y_\A)^*=\bar y_{\dot \A},~~~~(\bar y_{\dot\A})^*= y_{ \A},~~~~(z_\A)^*=\bar z_{\dot \A},~~~~(\bar z_{\dot\A})^*= z_{ \A},
\fe
such that $y\sigma_{dS}^\m \bar y$ is real and $(y\sigma_{dS}^\m\sigma_{dS}^\n  y)^*=\bar y\sigma_{dS}^\m\sigma_{dS}^\n \bar y$. We also define $(\sigma_{dS}^{\m\n})_{\A\B}=(\sigma_{dS}^{[\underline{\m}})_{\A}{}^{\dot\C}(\sigma_{dS}^{\underline{\n}]})_{\B\dot\C}$.


\begin{thebibliography}{}

%\cite{Strominger:2001pn}
\bibitem{Strominger:2001pn} 
  A.~Strominger,
  ``The dS/CFT correspondence,''
  JHEP {\bf 0110}, 034 (2001)
  [hep-th/0106113]; ibid ``Inflation and the dS/CFT correspondence,''
  JHEP {\bf 0111}, 049 (2001)
  [hep-th/0110087].

  %\cite{Witten:2001kn}
\bibitem{Witten:2001kn} 
  E.~Witten,
  ``Quantum gravity in de Sitter space,''
  hep-th/0106109.
  %%CITATION = HEP-TH/0106109;%%
  %412 citations counted in INSPIRE as of 08 Aug 2013

  
  
  %\cite{Maldacena:2002vr}
\bibitem{Maldacena:2002vr} 
  J.~M.~Maldacena,
  ``Non-Gaussian features of primordial fluctuations in single field inflationary models,''
  JHEP {\bf 0305}, 013 (2003)
  [astro-ph/0210603].
  %%CITATION = ASTRO-PH/0210603;%%
  %1016 citations counted in INSPIRE as of 08 Aug 2013

%\cite{Anninos:2011ui}
\bibitem{Anninos:2011ui}
D.~Anninos, T.~Hartman and A.~Strominger,
``Higher Spin Realization of the dS/CFT Correspondence,''
arXiv:1108.5735 [hep-th].
%%CITATION = ARXIV:1108.5735;%%<br /> 49 citations counted in INSPIRE as of 15 May 2013


%\cite{Vasiliev:1992av}
\bibitem{Vasiliev:1992av}
M.~A.~Vasiliev,
``More on Equations of Motion for Interacting Massless Fields of All Spins in (3+1)-Dimensions,''
Phys.\ Lett.\ B {\bf 285} (1992) 225.
%%CITATION = PHLTA,B285,225;%%<br /> 164 citations counted in INSPIRE as of 21 May 2013


%\cite{Vasiliev:1995dn}
\bibitem{Vasiliev:1995dn}
M.~A.~Vasiliev,
``Higher Spin Gauge Theories in Four-Dimensions, Three-Dimensions, and Two-Dimensions,''
Int.\ J.\ Mod.\ Phys.\ D {\bf 5} (1996) 763
[hep-th/9611024].
%%CITATION = HEP-TH/9611024;%%<br /> 189 citations counted in INSPIRE as of 21 May 2013


%\cite{Vasiliev:1999ba}
\bibitem{Vasiliev:1999ba}
M.~A.~Vasiliev,
``Higher Spin Gauge Theories: Star Product and AdS Space,''
In *Shifman, M.A. (ed.): The many faces of the superworld* 533-610
[hep-th/9910096].
%%CITATION = HEP-TH/9910096;%%<br /> 226 citations counted in INSPIRE as of 21 May 2013

%\cite{Kristiansson:2003xx}
\bibitem{Kristiansson:2003xx} 
  F.~Kristiansson and P.~Rajan,
  ``Scalar field corrections to AdS(4) gravity from higher spin gauge theory,''
  JHEP {\bf 0304}, 009 (2003)
  [hep-th/0303202].
  %%CITATION = HEP-TH/0303202;%%
  %18 citations counted in INSPIRE as of 13 Aug 2013
 
%\cite{Sezgin:2002ru}
\bibitem{Sezgin:2002ru} 
  E.~Sezgin and P.~Sundell,
  ``Analysis of higher spin field equations in four-dimensions,''
  JHEP {\bf 0207}, 055 (2002)
  [hep-th/0205132].
  %%CITATION = HEP-TH/0205132;%%
  %67 citations counted in INSPIRE as of 13 Aug 2013

%\cite{Vasiliev:1989yr}
\bibitem{Vasiliev:1989yr} 
  M.~A.~Vasiliev,
  ``Dynamics Of Massless Higher Spins In The Second Order In Curvatures,''
  Phys.\ Lett.\ B {\bf 238}, 305 (1990).
  %%CITATION = PHLTA,B238,305;%%
  %22 citations counted in INSPIRE as of 15 Aug 2013

%\cite{Vasiliev:2003ev}
\bibitem{Vasiliev:2003ev} 
  M.~A.~Vasiliev,
  ``Nonlinear equations for symmetric massless higher spin fields in (A)dS(d),''
  Phys.\ Lett.\ B {\bf 567}, 139 (2003)
  [hep-th/0304049].
  %%CITATION = HEP-TH/0304049;%%
  %249 citations counted in INSPIRE as of 09 Aug 2013

%\cite{Manvelyan:2005fp}
\bibitem{Manvelyan:2005fp} 
  R.~Manvelyan and W.~Ruhl,
  %``The Off-shell behaviour of propagators and the Goldstone field in higher spin gauge theory on AdS(d+1) space,''
  Nucl.\ Phys.\ B {\bf 717}, 3 (2005)
  [hep-th/0502123].
  %%CITATION = HEP-TH/0502123;%%
  %13 citations counted in INSPIRE as of 22 Sep 2013

%\cite{Bekaert:2005vh}
\bibitem{Bekaert:2005vh} 
  X.~Bekaert, S.~Cnockaert, C.~Iazeolla and M.~A.~Vasiliev,
  ``Nonlinear higher spin theories in various dimensions,''
  hep-th/0503128.
  %%CITATION = HEP-TH/0503128;%%
  %272 citations counted in INSPIRE as of 09 Aug 2013

%\cite{Giombi:2009wh}
\bibitem{Giombi:2009wh} 
  S.~Giombi and X.~Yin,
  ``Higher Spin Gauge Theory and Holography: The Three-Point Functions,''
  JHEP {\bf 1009}, 115 (2010)
  [arXiv:0912.3462 [hep-th]].
  %%CITATION = ARXIV:0912.3462;%%
  %107 citations counted in INSPIRE as of 19 Mar 2013
  
  %\cite{Giombi:2012ms}
\bibitem{Giombi:2012ms} 
  S.~Giombi and X.~Yin,
  ``The Higher Spin/Vector Model Duality,''
  arXiv:1208.4036 [hep-th].
  %%CITATION = ARXIV:1208.4036;%%
  %21 citations counted in INSPIRE as of 19 Mar 2013


%\cite{Giombi:2011kc}
\bibitem{Giombi:2011kc}
S.~Giombi, S.~Minwalla, S.~Prakash, S.~P.~Trivedi, S.~R.~Wadia and X.~Yin,
``Chern-Simons Theory with Vector Fermion Matter,''
Eur.\ Phys.\ J.\ C {\bf 72} (2012) 2112
[arXiv:1110.4386 [hep-th]].
%%CITATION = ARXIV:1110.4386;%%<br /> 45 citations counted in INSPIRE as of 16 May 2013


%\cite{Chang:2012kt}
\bibitem{Chang:2012kt} 
  C.~-M.~Chang, S.~Minwalla, T.~Sharma and X.~Yin,
  ``ABJ Triality: from Higher Spin Fields to Strings,''
  arXiv:1207.4485 [hep-th].
  %%CITATION = ARXIV:1207.4485;%%
  %29 citations counted in INSPIRE as of 18 Mar 2013
  
 


%\cite{Iazeolla:2007wt}
\bibitem{Iazeolla:2007wt}
C.~Iazeolla, E.~Sezgin and P.~Sundell,
``Real Forms of Complex Higher Spin Field Equations and New Exact Solutions,''
Nucl.\ Phys.\ B {\bf 791} (2008) 231
[arXiv:0706.2983 [hep-th]].
%%CITATION = ARXIV:0706.2983;%%<br /> 24 citations counted in INSPIRE as of 15 May 2013


%\cite{Aharony:2011jz}
\bibitem{Aharony:2011jz}
O.~Aharony, G.~Gur-Ari and R.~Yacoby,
``D=3 Bosonic Vector Models Coupled to Chern-Simons Gauge Theories,''
JHEP {\bf 1203} (2012) 037
[arXiv:1110.4382 [hep-th]].
%%CITATION = ARXIV:1110.4382;%%<br /> 35 citations counted in INSPIRE as of 15 May 2013


%\cite{Aharony:2012nh}
\bibitem{Aharony:2012nh}
O.~Aharony, G.~Gur-Ari and R.~Yacoby,
``Correlation Functions of Large $N$ Chern-Simons-Matter Theories and Bosonization in Three Dimensions,''
JHEP {\bf 1212} (2012) 028
[arXiv:1207.4593 [hep-th]].
%%CITATION = ARXIV:1207.4593;%%<br /> 17 citations counted in INSPIRE as of 16 May 2013


%\cite{GurAri:2012is}
\bibitem{GurAri:2012is}
G.~Gur-Ari and R.~Yacoby,
``Correlators of Large $N$ Fermionic Chern-Simons Vector Models,''
JHEP {\bf 1302} (2013) 150
[arXiv:1211.1866 [hep-th]].
%%CITATION = ARXIV:1211.1866;%%<br /> 3 citations counted in INSPIRE as of 16 May 2013

%\cite{Sezgin:2012ag}
\bibitem{Sezgin:2012ag} 
  E.~Sezgin and P.~Sundell,
  ``Supersymmetric Higher Spin Theories,''
  J.\ Phys.\ A {\bf 46}, 214022 (2013)
  [arXiv:1208.6019 [hep-th]].
  %%CITATION = ARXIV:1208.6019;%%
  %3 citations counted in INSPIRE as of 03 Jul 2013
  ].
  
%\cite{Banerjee:2013mca}
\bibitem{Banerjee:2013mca} 
  S.~Banerjee, A.~Belin, S.~Hellerman, A.~Lepage-Jutier, A.~Maloney, D.~Radicevic and S.~Shenker,
  ``Topology of Future Infinity in dS/CFT,''
  arXiv:1306.6629 [hep-th]. 
  %%CITATION = ARXIV:1306.6629;%%
  %2 citations counted in INSPIRE as of 07 Aug 2013

%\cite{Karch:2013oqa}
\bibitem{Karch:2013oqa} 
  A.~Karch and C.~F.~Uhlemann,
  ``Higher-spin realization of a dS static patch/cut-off CFT correspondence,''
  arXiv:1306.0582 [hep-th]. 
  %%CITATION = ARXIV:1306.0582;%%
  %1 citations counted in INSPIRE as of 07 Aug 2013

%\cite{Das:2013qea}
\bibitem{Das:2013qea} 
  D.~Das, S.~R.~Das and G.~Mandal,
  ``Double Trace Flows and Holographic RG in dS/CFT correspondence,''
  arXiv:1306.0336 [hep-th]. 
  %%CITATION = ARXIV:1306.0336;%%

%\cite{Anninos:2013rza}
\bibitem{Anninos:2013rza} 
  D.~Anninos, F.~Denef, G.~Konstantinidis and E.~Shaghoulian,
  ``Higher Spin de Sitter Holography from Functional Determinants,''
  arXiv:1305.6321 [hep-th]. 
  %%CITATION = ARXIV:1305.6321;%%
  %5 citations counted in INSPIRE as of 07 Aug 2013


%\cite{Jin:2013lqa}
\bibitem{Jin:2013lqa} 
  K.~Jin,
  ``Higher Spin Gravity and Exact Holography,''
  arXiv:1304.0258 [hep-th]. 
  %%CITATION = ARXIV:1304.0258;%%
  %1 citations counted in INSPIRE as of 07 Aug 2013

%\cite{Anninos:2012ft}
\bibitem{Anninos:2012ft} 
  D.~Anninos, F.~Denef and D.~Harlow,
  ``The Wave Function of Vasiliev's Universe - A Few Slices Thereof,''
  arXiv:1207.5517 [hep-th]. 
  %%CITATION = ARXIV:1207.5517;%%
  %13 citations counted in INSPIRE as of 07 Aug 2013

%\cite{Das:2012dt}
\bibitem{Das:2012dt} 
  D.~Das, S.~R.~Das, A.~Jevicki and Q.~Ye,
  ``Bi-local Construction of Sp(2N)/dS Higher Spin Correspondence,''
  JHEP {\bf 1301}, 107 (2013)
  [arXiv:1205.5776 [hep-th]]. 
  %%CITATION = ARXIV:1205.5776;%%
  %13 citations counted in INSPIRE as of 07 Aug 2013
%8 citations counted in INSPIRE as of 07 Aug 2013

%\cite{Anninos:2012qw}
\bibitem{Anninos:2012qw} 
  D.~Anninos,
  ``De Sitter Musings,''
  Int.\ J.\ Mod.\ Phys.\ A {\bf 27}, 1230013 (2012)
  [arXiv:1205.3855 [hep-th]]. 
  %%CITATION = ARXIV:1205.3855;%%
  %12 citations counted in INSPIRE as of 07 Aug 2013

%\cite{Ng:2012xp}
\bibitem{Ng:2012xp} 
  G.~S.~Ng and A.~Strominger,
  ``State/Operator Correspondence in Higher-Spin dS/CFT,''
  Class.\ Quant.\ Grav.\  {\bf 30}, 104002 (2013)
  [arXiv:1204.1057 [hep-th]].
  %%CITATION = ARXIV:1204.1057;%%
  %20 citations counted in INSPIRE as of 07 Aug 2013
%\cite{Vanchurin:2012xm}

%\cite{Mkrtchian:1981bb}
\bibitem{Mkrtchian:1981bb} 
  R.~L.~Mkrtchian,
  ``The Equivalence Of Sp(2n) And So(2n) Gauge Theories,''
  Phys.\ Lett.\ B {\bf 105}, 174 (1981).
  %%CITATION = PHLTA,B105,174;%%
  %20 citations counted in INSPIRE as of 09 Aug 2013

 \end{thebibliography}
\end{document}